\documentclass[11pt]{article}
\usepackage{epsfig}
\usepackage{amsfonts}

\def\laq{~\raise 0.4ex\hbox{$<$}\kern -0.8em\lower 0.62
ex\hbox{$\sim$}~}
\def\gaq{~\raise 0.4ex\hbox{$>$}\kern -0.7em\lower 0.62
ex\hbox{$\sim$}~}

\def\beq{\begin{equation}}
\def\eeq{\end{equation}}
\def\bea{\begin{eqnarray}}
\def\eea{\end{eqnarray}}

\def \ti {\widetilde}

\def \La {\Lambda}

\def \b {\beta}
\def \a {\alpha}

\def \ga {\gamma}

\def \da {\delta}
\def \ep {\epsilon}

\def \noi {\noindent}

\def \Mp {M_{\rm P}}

\begin{document}
\begin{titlepage}

\begin{flushright}
BA-TH/06-555\\
CERN-PH-TH/2006-031\\
hep-th/0611227
\end{flushright}

\vspace{0.5 cm}

\begin{center}

\Huge{A new scale in the sky}

\vspace{1cm}

\large{M. Gasperini}\footnote{This paper was written while the author was affiliated to CERN as Scientific Associated at the Theory Unit, Physics Department.}

\bigskip
\normalsize

{\sl Dipartimento di Fisica, Universit\`a di Bari, \\
Via G. Amendola 173, 70126 Bari, Italy}, 

{\sl Istituto Nazionale di Fisica Nucleare, Sezione di Bari, 
Bari, Italy}

and 

{\sl CERN, Theory Unit, Physics Department, \\ CH-1211 Geneva 23, Switzerland} 

\vspace{0.5cm}

\begin{abstract}
\noi
The existence of a new ultraviolet scale $\La= g\Mp$ for effective theories with gravity and $U(1)$ gauge fields has recently been conjectured as a possible criterion for distinguishing parts of the swampland from the string landscape. Here we discuss a possible phenomenological signature of this scale, for electromagnetic fields, in astrophysical observations.
\end{abstract}
\end{center}

\smallskip
\begin{center}
---------------------------------------------\\
\vspace {5 mm}
{\em Essay written for the {2006 
Awards for Essays on Gravitation}}\\
{\em (Gravity Research Foundation, Wellesley Hills, MA, 02481-0004,
USA)\\
and awarded with ``Honorable Mention"}
%\end{center}

\vspace {8 mm}
%\bigskip
To appear in {\bf Int. J. Mod. Phys. D} (December 2006, Special Issue)
\end{center}

\vfill
\begin{flushleft}
CERN-PH-TH/2006-031\\
February 2006
\end{flushleft}

\end{titlepage}

\newpage
\parskip 0.2cm

The presence of a new ultraviolet scale $\La=g\Mp$ for effective theories with gravity and $U(1)$ gauge fields, with coupling $g$, has recently been conjectured in the first {\tt hep-th} paper of the year \cite{1}. Theoretical evidence for this conjecture has been presented, mainly based on the properties of known (heterotic) string theory backgrounds, and it has been argued that this conjecture could be related to a new criterion for distinguishing parts of the ``swampland" \cite{2} from the string ``landscape". 

According to \cite{1}, the conjectured scale  $\La=g\Mp$ is hidden, and completely unexpected, in the context of the effective field theory; here we will suggest that, rather surprisingly, there could be a hint at such a scale in present astrophysical observations. We are referring, in particular, to the ``electro-gravity" scale  $\La \sim \sqrt{\a_{\rm em}/ G_N} \sim e \Mp$. The physical relevance of such a scale might emerge from observed regularities  concerning the mass, the angular momentum, and the magnetic moment of various classes of astrophysical structures. Such regularities were noticed long ago but, up to now, have not been unambiguously explained in the context of standard (effective) models of gravity and electromagnetic interactions. Here we discuss the possibility that they be related to the general weakness of gravity with respect to gauge forces, according to the conjecture presented in \cite{1}. 

We start by recalling an empirical relation which seems to connect the mass $M$  and the angular momentum $J$ of many classes of astronomical objects, ranging from planets to superclusters. The relation was originally proposed in the form \cite{3,3a} $J \propto M^2$, and later extended to a generic slope $\da$ as follows \cite{4,5}:
\beq
J= p \,M^\da,
\label{1}
\eeq
where $p$ is some ``universal" phenomenological parameter to be determined by observations. Detailed compilations of data have shown the presence of a considerable scattering of points in the $J-M$ plane within the various astrophysical classes, but the overall fit of the plots of $\log J$ versus $\log M$, for all classes of objects, has confirmed that Eq. (\ref{1}) may be representative of the general trend of the data. In that case, the slope $\da$ has been determined to be $\da=1.98 \pm 0.04$ according to a first compilation \cite{4}, and 
$\da=1.94 \pm 0.09$ in a subsequent, larger compilation of data running over about $25$ powers of ten in the mass \cite{5} (see Fig. 1, left panel). Both results are compatible with a power $\da=2$. The experimental value of the (dimensional) coefficient $p$ is strongly correlated to the slope $\da$, and is thus affected by a large uncertainty. If we fix $\da=2$, however, the corresponding mean value of $p$ is very similar in the two cases, as one obtains (in cgs units) 
$p \simeq 8 \times 10^{-16}\, {\rm g^{-1} cm^2 s^{-1}}$ according to \cite{4}, and $p \simeq 3 \times 10^{-15}\, {\rm g^{-1} cm^2 s^{-1}}$ from the data of \cite{5}.

It must be remarked, at this point, that the case with slope $\da=2$ is qualitatively different from the generic case $\da\not=2$, because for $\da=2$ the dimensions of $p$ are such that we can express $p$ in terms of only two (dimensional) fundamental parameters, the  Newton constant $G_N$ and the ligth velocity $c$ (indeed, $[p]=[G_N c^{-1}]$). If, instead,  $\da\not=2$, then the definition of $p$ requires,  besides $G_N$ and $c$, another parameter with mass dimensions, since $[p]=[G_NM^{2-\da} c^{-1}]$, and the possibility of a universal parametrization is lost. We are thus lead to the conjecture that the small deviations of $\da$ from $2$ may represent ``systematic" perturbations (due to local-physics effects) of some underlying, ``scale-free" fundamental relation, that can be empirically parametrized by setting $\da=2$ and 
\beq
p^{-1}= \b\, G_N^{-1}c, ~~~~~~~~~~~~~~ \da=2,
\label{2}
\eeq
where $\b$ is a dimensionless number. With this assumption one  obtains from the data a mean value of $p$ ranging from $\b \simeq 3 \times 10^{-3}$, according to $\cite{4}$, to $\b \simeq 7 \times 10^{-4}$, from the results of \cite{5}. 

A second empirical relation concerns the approximated proportionality between the angular momentum and the magnetic moment ${\cal M}$ of astronomical bodies. By setting
\beq
J= q \, {\cal M}^\ep
\label{4}
\eeq
one then finds that for $\ep=1$ the numerical value of the (dimensional) parameter $q$ is of the same order of magnitude for many stars and planets. The recurrence of this direct proportionality coefficient (sometimes referred  as the ``magnetic Bode's law" \cite{6a}) was pointed out  long ago by Blackett \cite{7} for the Earth, the Sun and the star $78$ Virginis, and was later extended to other planets, white dwarfs, and pulsars in \cite{7a,8}. After Blackett, new and direct tests of this relation have been proposed also for macroscopic spinning bodies in the laboratory \cite{7a,8,9a}, and the required sensitivity for these experiments seems now to be within the reach of modern technology. 
 
\begin{figure}[t]
  \vspace{-1cm}
\centerline{\includegraphics[width=65mm]{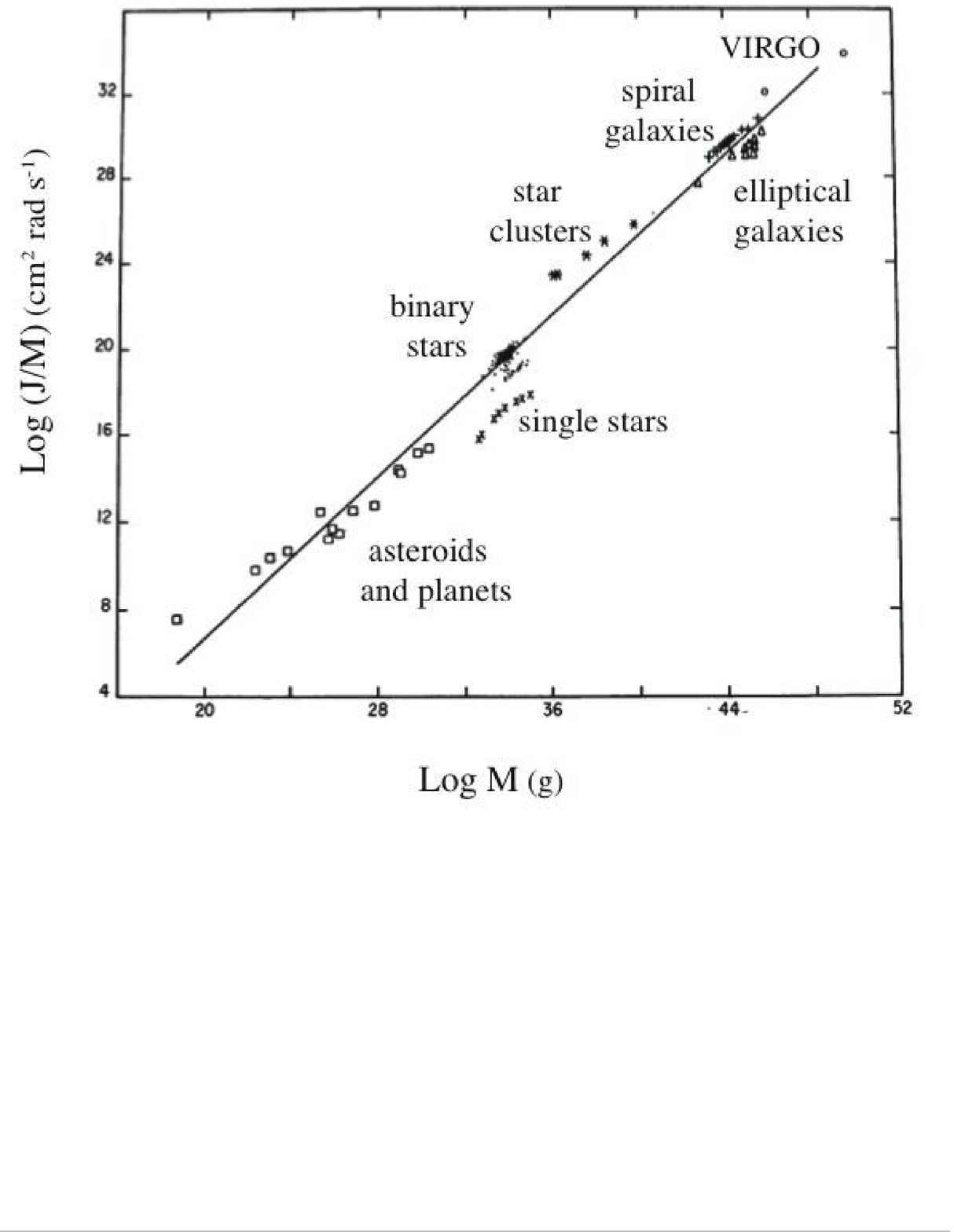}
 ~~~\includegraphics [width=79mm]{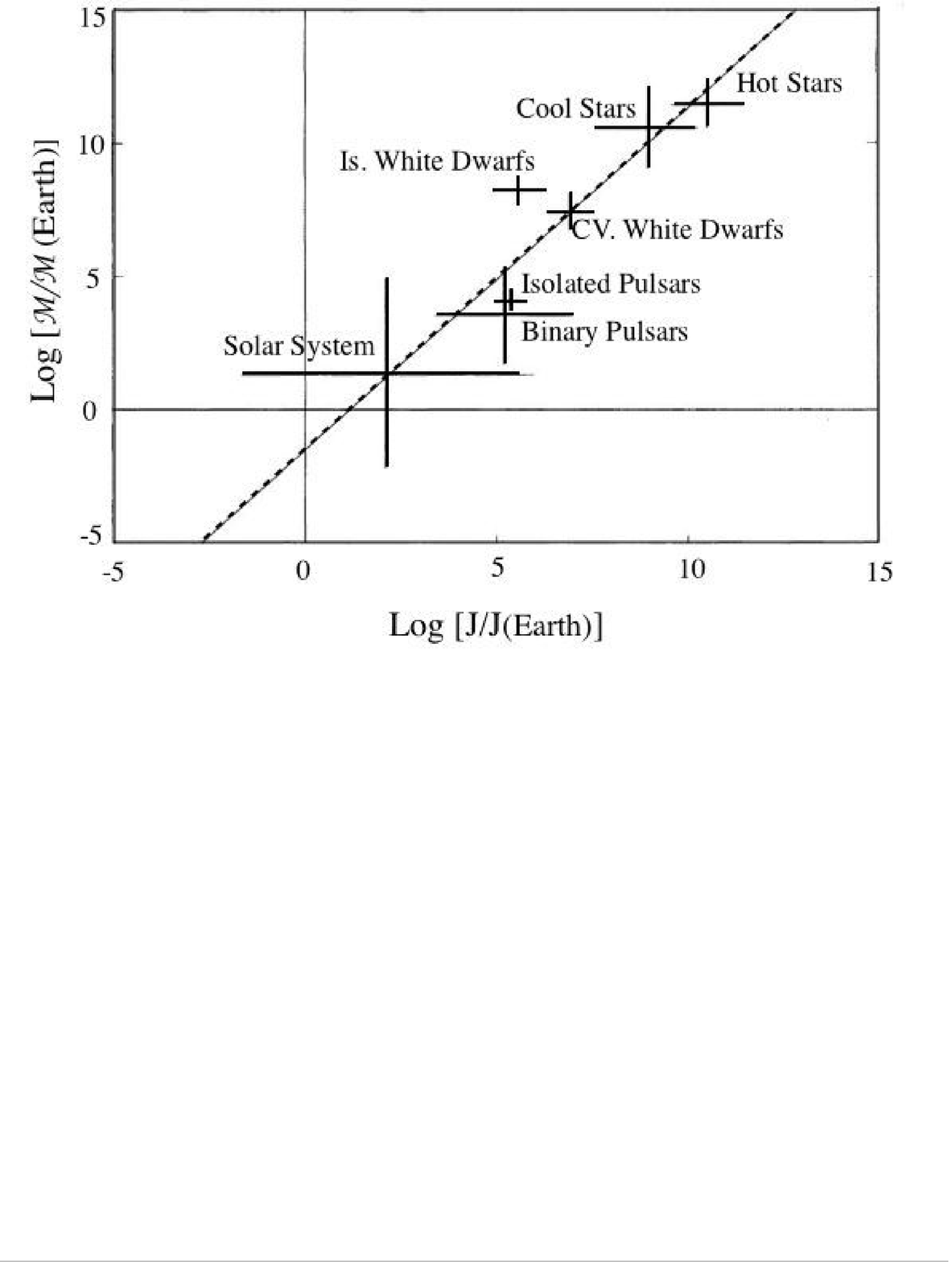}}
 \vspace{-3cm}
\caption{\sl The left panel shows the logarithmic plot of the angular momentum per unit mass versus the mass, for a large variety of astrophysical systems (adapted from Ref. \cite{5}). The right panel shows a similar magnetic moment/angular momentum diagram, where the bars denote the range of values of $J$ and $\cal M$ spanned within the various sub-sambles (adapted from Ref. \cite{8a}).} 
\label{f1}
\end{figure}

It is unclear, at present, whether such an empirical correlation may retain universal validity when extrapolated to include other (and larger) classes of astronomical objects (the analyses are also complicated, in general, by the possible variability in time of the magnetic fields, by the presence of toroidal and polar magnetic components, \dots). A recent data analysis \cite{8a}, spanning a range of about $20$ orders of magnitude in both $J$ and ${\cal M}$, has indeed confirmed the trend suggested in \cite{7a,8} for the planetary sector, and extended it to a large compilation of stellar objects, in the sense that the plots of $\log {\cal M}$ versus $\log J$, for the ``centers of gravity" of the data of the various sub-samples, have been fitted by an overall slope $\ep^{-1}=1.294 \pm 0.318$, which is marginally compatible with $\ep=1$ (see Fig. \ref{f1}, right panel). However, it  has been shown that there are significant deviations from this mean result inside the sub-samples of data corresponding to different astrophysical classes (deviations from a linear $J-{\cal M}$ correlation are most evident for hot stars, isolated white dwarfs, and isolated pulsars \cite{8a,8b,8c}). 

The statistical significance of the proposed correlation fitting the complete set of data, averaged over all sub-samples,  is likely to remain controversial for a long time, as also noted in \cite{8b}. Leaving this problem to a forthcoming, more detailed analysis \cite{8d}, in this paper we will take seriously the results of \cite{8a}, and we shall concentrate the following discussion on the case $\ep=1$ which is (at least marginally) compatible with the data, and which is dimensionally preferred, as before, in the sense of allowing a scale-free parametrization of $q$ in terms of $c$ and $G_N$ only. When  $\ep\not=1$ the definition of $q$ requires indeed an extra length parameter, as $[q]= [c^{3-2\ep} L^{2-2\ep} G_N^{-1+\ep/2}]$; with  $\ep=1$ we can instead adopt a universal parametization, and set
\beq
q^{-2}= \ga \, G_N c^{-2}, ~~~~~~~~~~~~~ \ep=1, 
\label{5}
\eeq
where $\ga$ is a dimensionless coefficient. Using the data reported in  \cite{8a}, and the mean value of $q$ associated to the slope $\ep=1$,  one then finds $\ga \simeq 9 \times 10^{-3}$. The result is very close to the value of $\ga$ one would obtain from the old analyses of \cite{7a,8},  mainly restricted to the planetary sector. 

Now we come to the new and central point of this paper, namely to the observation that the experimental values of the two dimensionless coefficients $\b$ and $\ga$ turn out to be not only very close to each other, within observational uncertainties, but also very close to the numerical value of the fine-structure constant, $\a_{\rm em} \simeq 7.3 \times 10^{-3}$. Barring an accidental numerical coincidence, we are thus led to our central assumption, namely to the speculation that the two coefficients $\b$ and $\ga$ actually represent {\em the same} physical parameter (modulo numerical factors of order $1$), and that this parameter actually corresponds to the fine-structure constant:
\beq
\b \simeq \ga \simeq \a_{\rm em} = e^2/\hbar c \simeq 1/137 \simeq 7.3 
\times 10^{-3}.
\label{7}
\eeq

This simple assumption, suggested by the astrophysical data, leads immediately to the disclosure of a direct connection between the new cut-off scales conjectured in \cite{1} and the two empirical laws (\ref{1}), (\ref{4}). According to Eqs. (\ref{2}), (\ref{7}), in fact, the parameter $p$ determines an energy scale $\La$ given (in natural units $\hbar=c=1$) by 
\beq
p^{-1/2} \sim \La = \sqrt{\a_{\rm em}}\, \Mp = e\, \Mp,
\label{8}
\eeq
which is exactly the new UV scale suggested in \cite{1}. According to Eqs. (\ref{5}), (\ref{7}), the parameter $q$ determines instead the ``dual" magnetic scale 
\beq
q \sim \ti \La =  {\Mp\over  \sqrt{\a_{\rm em}}} = {\Mp\over e}= g_{\rm mag} \Mp,
\label{9}
\eeq
also suggested in \cite{1}. It seems remarkable that these two related scales emerge {\em separately} and quite {\em independently} from two different (and apparently unrelated) empirical laws. 
One might see in this result a signal supporting the conjectures of \cite{1}.  

It should be pointed out, however, that the numerical value of $\a_{\rm em} \simeq 1/137$ is not very different from  a realistic asymptotic value of the square of the GUT gauge-coupling parameter, $\a_{\rm em} \sim \a_{\rm GUT}^2$. It is a curious coincidence that the replacement of $\a_{\rm em}$ with $\a_{\rm GUT}^2$ in Eq. (\ref{8}) would lead to extract from the data another energy scale, $\La= \a_{\rm GUT} \Mp$, which also may have the meaning of UV cut-off scale (numerically similar but conceptually very different from $e \Mp$), as recently stressed 
on the grounds of the principle of asymptotic saturation of all couplings, induced by loop corrections \cite{12a}. The additional relation (\ref{9}), and the presence of macroscopic magnetic fields in the relevant macroscopic observations, however, would seem to point at the electro-gravity scales (\ref{8}), (\ref{9}) as a possibly more appropriate interpretation of the data. 

It should be mentioned that the near coincidence of $\b$ and  $\sqrt{\a_{\rm em}}$ was previously noticed in \cite{4}, without any reference, however, to the empirical law concerning magnetic moments. Also, a theoretical interpretation of the parameter $q$, as a possible imprint of early geometric spin-torsion interactions at the epoch of structure formation, was attempted in \cite{9}, providing the result $q \sim (\a_{\rm em}\, G_n)^{-1/2}$, in agreement with Eq. (\ref{9})  (leaving unexplained, however, the mass-angular momentum relation) . The interpretation of $\La$, $ \ti \La$ as effective cut-off scales, following the conjectures of \cite{1}, seems instead to connect two different empirical relations, and to predict a limiting scale for the angular momentum and the magnetic moment of macroscopic bodies of mass $M$, fixing a lower bound on the {\em maximal} allowed values of  $J$ and ${\cal M}$: 
\beq
J \gaq \left(M\over e \Mp\right)^2, ~~~~~~~~~~~~~~~
{\cal M} \gaq {eJ\over \Mp}
\label{10}
\eeq
(in natural units, from the combination of Eqs. (\ref{1}), (\ref{8}) and  Eqs. (\ref{4}), (\ref{9}), respectively). 

As remarked in \cite{1} for other restrictions implied by the UV scale $\La$, this {\em does not} mean that it is impossible to have states with $J$ and ${\cal M}$ smaller than the bound (\ref{10}), but it means that there must exist states satisfying the above inequalities. The lesson we could learn, eventually, from the data is that the astrophysical structures, in the absence of perturbations, try to accumulate along a ``preferred" configuration corresponding to the state with the highest values of  $J$ and ${\cal M}$ compatible with the saturation of  the above bound, required for a Universe safely placed inside the string landscape \cite{1}.

\section*{Acknowledgements}
It is a pleasure to thank Gabriele Veneziano for clarifying discussions, and Dharamvir Ahulawalia-Khalilova for pointing out some interesting papers to my attention. 

%\newpage


\begin{thebibliography}{99}
\newcommand{\bb}{\bibitem}

\bb{1}N. Arkani-Hamed, L. Motl, A. Nicolis and C. Vafa, hep-th/0601001. 

\bb{2}C. Vafa, hep-th/0509212. 

\bibitem{3}P. Brosche, Z. Astrophys. {\bf 57}, 143 (1963); Astrophys. Space Sci. {\bf 29}, L7 (1974). 

\bb{3a}P. Wesson, Astron. Astrophys. {\bf 80}, 296 (1979).

\bb{4}P. Wesson,  Phys. Rev.  {\bf D23}, 1730 (1981).

\bb{5}L. Carrasco, M. Roth and A. Serrano, Astr. Astrophysics {\bf 106}, 89 (1982). 

\bb{6a}C. T. Russel, Nature {\bf 272}, 147 (1978). 

\bb{7}P. M. S. Blackett, Nature {\bf 159}, 658 (1947).

\bb{7a}D. V. Ahluwalia and T. -Y. Wu, Lett. Nuovo Cimento {\bf 23}, 406 (1978).

\bb{8}S. P. Sirag, Nature {\bf 278}, 535 (1979).

\bb{9a}M. Surdin, J. Franklin Institute {\bf 303}, 493 (1977).

\bb{8a}C. N. Arge, D. J. Mullan and A. Z. Dolginov, Astrophys. J. {\bf 443}, 795 (1995).

\bb{8b}S. Baliunas, D. Sokoloff and W. Soon, Astrophys. J. {\bf 457}, L99 (1996).

\bb{8c}B. V. Vasiliev, astro-ph/0002048. 

\bb{8d}M. Gasperini et al., in preparation. 

\bb{12a}G. Veneziano, JHEP {\bf 0206}, 051 (2002). 

\bb{9}V. De Sabbata and M. Gasperini, in {\em The origin and Evolution of Galaxies} \\ (World Scientific, Singapore, 1982), p. 181; 
Acta Cosmologica {\bf 12}, 79 (1983). 

\end{thebibliography}
\end{document}